# Establishing digital molecular communications in blood vessels

Luca Felicetti, Mauro Femminella, Gianluca Reali
Department of Electronic and Information Engineering, University of Perugia
Perugia, 06125, Italy
{luca.felicetti, mauro.femminella, gianluca.reali}@diei.unipg.it

*Abstract*—In this paper we propose a solution for transmitting digital information within the cardiocirculatory system. In particular, we make use of a channel delivering burst of molecules, emitted by mobile transmitters, which diffuse in the blood towards fixed receivers, that are attached to the vessel walls. This communication scheme has been inspired by the real signaling between platelets and endothelial cells, the behavior of which has been investigated experimentally. We thus believe that our proposal can be successfully deployed in living bodies. On the basis of the results achieved through simulations on the communication system capabilities, we propose a simple but effective receiver scheme, and we outline the future research directions.

## I. INTRODUCTION

Nanoscale communications is an emergent research topic with potential innovative applications in many fields [1][2], including medicine [3]. Even if research on the design of biocompatible nanomachines is an active research topic for many years [4], organizing effective and reliable communications among them is still challenging. In order to design nanomachines able to exploit the communications potentials of nanoscale environments, it is necessary to identify the basic communications mechanisms happening at the nanoscale and the relevant parameters.

A special environment for implementing nanoscale communications is the cardiocirculatory system. In fact, the continuous monitoring of the concentrations of specific parameters in the blood could greatly help doctors to detect potential critical situations for patients. With the availability of such information, it could be possible to trigger not only the early administration of the necessary therapies in standard ways, but also the release of specific medicines by nano-actuators previously deployed in the human body. Since the blood requires about one minute to complete the large and small circulation, it is an extraordinary medium to quickly carry information to different parts of the body. Thus, understanding the communications mechanisms in the blood is of great interest in the field on nanoscale communications.

In this work, we model molecular communications inside blood vessels between mobile transmitters and fixed receivers, which are sticky to the vessel walls. In more detail, we present a study on the response to a pulse of carriers released by a mobile transmitter, measured on a number of receivers located along the vessel wall. This study has been done by using the BiNS simulator [16], tuned by using the results of in-vitro experiments, which in the last version is able to simulate bounded spaces, employing also to parallel simulation techniques [17] to handle effectively the huge amount of derived collisions between the involved mobile and fixed entities. On the basis of the obtained results, we propose a simple but effective scheme for implementing a biocompatible digital receiver, able to auto-synchronize and decode the information received by means of repeated pulses of information molecules.

The work is organized as follows. In section II, we illustrate the work background, and review related works in the field. In section III we illustrate the model for the receiver. In section IV, we present the numerical results obtained by simulations, discussing the design choices and selecting the appropriate configuration for the receiver. Finally, in section V, we draw our conclusion and outline future research directions.

## II. BACKGROUND AND RELATED WORK

### A. Biological background for signaling in bloodstream

In order to model the communications between nanomachines inside the bloodstream, we were inspired by the well established signaling system between the platelets (mobile transmitters) and endothelial cells (fixed receivers), based on the release of specific proteins known as CD40L [21], which are able to bind to the relevant receptors (CD40) available on the surface of the receivers. In fact, in normal condition, this mechanism regulates the response to injury inside the vessels, allowing to platelets to repair the vessel damage, activate the endothelium so that to recruit monocytes in the injury site to fight against eventual infections. However, in pathological conditions, it is also recognized as the initial trigger to the development of atherothrombotic diseases [20][22].

### B. Background on the propagation models in bloodstream

The communication environment of the bloodstream is particularly complex, for both mobile transmitters and released information molecules. In fact, it consists of three components:

1. particle diffusion, which can be modeled by the theory

of the Brownian motion [7];

2. positive drift due to the pressure of the bloodstream, which depends on the position of the particle with respect to the axis of the vessels (parabolic profile, according to the Poiseuille or Casson laws [13]);

3. movement resulting from collisions between particles.

Whilst there are a number of theoretical models taking into account diffusion [7] and diffusion with drift for nanoparticles (see [13] and references therein), or communication by contact the analysis of all the three components together has been published with some details only recently [12]. In particular, the study in [12] shows that the influence of the red blood cells within the bloodstream is dominant, since they are the most numerous particles present in the blood vessels, and their size is (slightly) inferior only to that of the white blood cells. Thus, the resulting effect is that smaller particles (not only nanoparticles as shown in [12], but also platelets [14]) are pushed towards the vessel walls.

Finally, this work is different from [25], in which the authors considers mobile nanomachines and model collisions, but information transfer occurs by using electrochemical communication upon physical contact with each other.

*C. Related work on molecolar communications*

Works dealing with molecolar communications propose to use nanoparticles as information signal. The signal propagation mechanism consists of their diffusion in the fluid medium, usually modeled as a Brownian motion. The paper [6] presents mathematical models of transmitter, channel, and receiver. Nodes are assumed to be fixed. The authors evaluate the end-to-end gain and propagation delay as a function of some environmental parameters. Information is transferred by modulating the concentration of the particles emitted by the transmitter. In the same scenario, the works [9] models the noise sources affecting the diffusion-based molecular communications, using the number of bound receptor (i.e. forming a receptor-ligand complex) as the input signal at the receiver. All these works assume a transmission medium in which propagation is due to diffusion only, which does not apply to our scenario. In addition, the output reflects the current status of the system (i.e. the number of complexes).

A slightly different scenario is depicted in [10], where the transmitter sits within a fluid and emits a series of identical molecules, which disperse in the fluid by a Brownian motion and are absorbed by a receiver capable of measure their arrival times. The information is encoded in the release time of molecules as in a pulse position modulation. In [5] the authors propose a slightly different scenario, in which both nanomachines and emitted molecules propagate through a fluid medium, with a drift velocity and Brownian motion component. They analyze a model of a communication system based on the release of either one or two molecules into the medium. In [11], the same authors show that the additive inverse Gaussian noise channel is an appropriate model for molecular communications in fluid media with drift. They also derived upper and lower bounds of channel capacity and propose a maximum likelihood receiver. Although these scenarios are similar to the one of this paper, the transmission media are modeled without taking into account the massive contribution due to collisions with a large moving cells, as in [12]. Thus, their models cannot be easily applied to the bloodstream scenario.

III. THE COMMUNICATIONS SYSTEM MODEL

The block scheme of the end-to-end model is depicted in Fig. 1, by highlighting some details of the transmitter and the receiver. The transmitter encodes the received stimulus in a train of pulses, each one composed of $B$ molecules. These pulses are spaced in time each other by a time $\tau$. A bit 0 is encoded with no release of molecules, thus we are using an on-off keying technique. The first pulse is used to allow the receiver synchronizing with the transmitter. It is followed by a fixed number of $N$ pulses to encode the stimulus, i.e. the information. In this regard, we consider different choices to encode the information to be transmitted. The first option is to map each bit with a symbol $b_k$. This means that a train of $N$ pulses $b_k$ will encode a sequence of $N$ bits. Another choice is to encode a single group of $M$ information bits (symbols) with $K$ pulses $b_k$, with $K>M$. In more detail, the idea to be explored is to perform a sort of baseband CDMA, using trains of pulses which are uncorrelated or exhibit very low values for the correlation function. Even if this second option is more robust of the first one, a possible drawback could be associated to the timescales associated to the channel propagation, especially when considering mobile transmitters and fixed receivers, or vice versa, or in general when the speed of transmitter and receiver are different. In fact, in the considered environment, differently from classical communications using electromagnetic waves, the speed of transmitter and/or receiver is comparable with that of the signal. Since, in order to have a successful communication also in the case of molecular communications, the pulse period should be larger than the channel delay spread [23], the time necessary to transmit the whole train of pulses could be so large that the transmitter could be out range of the receiver (or vice versa). Thus an important design parameter will the length of the pulse train, in order to allow successful communications.

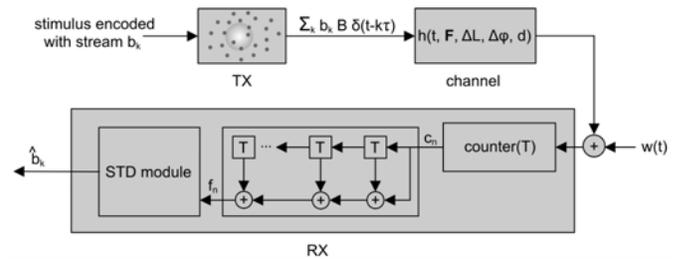

Fig. 1. Block scheme of the end-to-end trasmission chain.

As for the receiver, it is composed by a number of modules. The received signal consists of the number of nanoparticles (ligands) that get in contact with the surface of the receiver, which is populated by a number of receivers, compliant with the specific ligand. The module counter (T) counts the number of nanoparticles *assimilated* during the last time window of duration $T$, that is the complexes able to transmit a signal towards the nucleus of the receiving cells. Assimilation can imply or not the internalization of the complex ligand-receptor [19]. At the expiration $t_n$ of the *n*-th observation time, the counter is reset and it makes available a new reading $c_n$:

$$c_n = \{\text{\# of assimilations in } [t_n - T, t_n]\} \quad (1)$$

These measures feed a digital, finite impulse response (FIR) filter, acting as a sliding window of duration $(P+1) \times T$, being $P$ the number of delay lines of the filter. The output of this filter, $f_n$, is the input of the Synchronization and Threshold Detection (STD) module, with

$$f_n = \sum_{i=0}^{P} c_{n-i} \quad (2)$$

The STD module has two functions, described by the pseudo-code in Fig. 2: (i) synchronizing the receiver with a new train of pulses, and (ii) decoding the information through a decision threshold mechanism. When it is not synchronized, the STD observes the input signal every $T$ seconds until it detects a value of $f_n > T_h$. Since this time, it assumes to be synchronized with the transmitter, and reads the value of $f_n$ for $N$ times, each one spaced in time of $\tau=(P+1) \times T$ seconds, so that to take into account the delay introduced by the channel after the pulse transmission. Clearly, more sophisticated synchronization and encoding schemes are possible, which are part of our future research. The value of $\tau$ has to be determined as the maximum between the channel dispersion time and the time needed to allow transferring chemical information from the ligand-receptor complex to the receiving cell nucleus. Thus, the proposed receiver model resembles the functioning of a cell engineered to estimate the value of a stream of $N$ digital pulses after being activated (i.e. time synchronized with the transmitter) by the first synchronization pulse. The readers should bear in mind that the proposed scheme is applicable to both fixed and mobile receivers.

The distortion effect of the cannel is quite complex, and depends on a number of factors, including the set of characteristics of the blood flow, referred to as **F** in Fig. 1 (concentration and size of blood cells, pressure, temperature, etc.), and the displacement of the transmitter with respect to the considered receiver on the vessel wall, as illustrated in Fig. 3. The effect of the channel has been evaluated by simulation.

```
while forever do
   /* Idle, search for alignment */
   while (f_n < Th) do
      wait for T seconds
   done
   /* Now aligned, receiving N bits */
   for i:=1,..,N do
      wait for (P+1)×T seconds
      if (f_n < Th) then
         b_i:=0
      else
         b_i:=1
      end
   done
   /* Leave an interval to free system memory */
   wait for (P+1)×T seconds
done
```

Fig. 2. Pseudo-code illustrating STD functions.

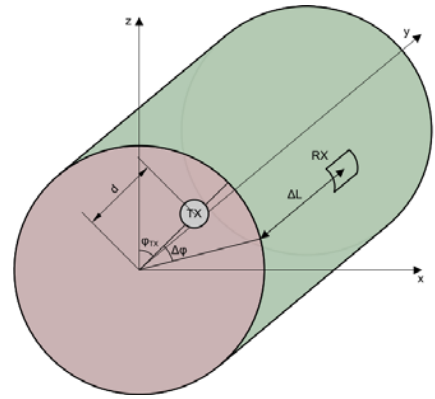

Fig. 3. Relative displacement of the transmitter with respect to a receiver.

In summary, the proposed communication scheme resembles that of a baseband digital communication system, since the transmitter transmits digital pulses, and the receiver has to reconstruct the binary stream in the presence of channel distortion.

IV. NUMERICAL RESULTS

In this section, we first analyze the effect introduced by the channel, and then evaluate the capability of the proposed receiver of estimating the transmitted signal. If the number of carriers received by the single receiver is much lower than the number of receptors, from our results the channel appears to be linear and stationary. Thus, we show the transmission of a single impulse (approximated through a short burst of carriers) to both analyze the impulse response of the channel and derive the relevant values needed to correctly configure the receiver. The simulation parameters are reported in TABLE I. As mentioned in section II.A, we have used the platelet experimental model, to simulate transmitter nodes, of endothelial cells as receiver nodes, and sCD40L as carriers. Nevertheless, the proposed scheme is general, and can be used also for analyzing different types molecular communications.

As for the mobile node used as transmitter, at the time of transmission its center has cylindrical coordinates equal to $\mathbf{X_0}=(\phi, d, L)=(\pi/4, i\times\Delta d, 0)$, where $i=0,..,5$ and $\Delta d$=5.425 µm, that is from the cylinder axis ($i$=0) up to a position very close to the vessel wall ($i$=5). The receivers form an annulus of 13 cells, deployed for values of $\Delta L$ in the range -0.4÷ 0.95 mm.

TABLE I. SIMULATION PARAMETERS

| General parameters | | Receivers cells (endothelial cells) | |
|---|---|---|---|
| Vessel length | 1.35 mm | Cell side | 15 μm |
| Vessel radius | 30 μm | # CD40 receptors | 1000 |
| Mean flow velocity | 0.5 mm/s | Receptors radius | 4 nm |
| Viscosity | 1.3 mPa×s | **White blood cells (WBC)** | |
| Temperature | 310° K | Concentration | $4 \times 10^3$ mm$^{-3}$ |
| **Transmitter cells (platelets)** | | # CD40 receptors | 2000 |
| Concentration | $2 \times 10^5$ mm$^{-3}$ | Radius | 5 μm |
| Radius | 1 μm | **Red blood cells (RBC)** | |
| # of receptors | 1000 | Concentration | $4 \times 10^6$ mm$^{-3}$ |
| Receptors radius | 4 nm | Radius | 3.5 μm |
| Burst size | 3000 | **Simulation parameter** | |
| sCD40L radius | 1.75 nm | Time step | 5 μs |

Fig. 4 shows the number of assimilated carriers by fixed receivers after 8 seconds of simulated time The position of the transmitter $\mathbf{X_0}$, at the transmission instant, is indicated by a red asterisk. Subfigures (a) and (b) are relevant to transmitter positions close to the vessel walls (the most likely, as reported in [14]), whereas Fig. 4.(c) shows the reception map when the transmitter is located over the longitudinal axis of the vessel. By analyzing Fig. 4.(a) and Fig. 4.(b), it emerges that the most significant effect due to a large distance between the transmitter and the vessel wall is the extension of the signal footprint in the direction of the longitudinal axis. It means that a higher number of cells are able to receive the signal, which thus results to be more attenuated, analogously to a highly directional antenna footprint where the antenna height is increased. The main difference is that the maximum value assumed by the total number of assimilated carriers is bit higher in Fig. 4.(a), and that for a position of the transmitter very close to the vessel wall, there are a small number of cells behind the transmitter able to receive the signal. This effect is due the bounce back of the carriers on the red blood cells or the transmitter itself. As for Fig. 4.(c), the cells able to receive the signal are less grouped. They are distributed over the whole $\phi$ range, without directionality in the footprint, and exhibit a symmetric cell coverage. In addition, the coordinates of the cells able to receive the signal are shifted towards positive values of the longitudinal axis. Fig. 4.(a) and Fig. 4.(c) show the coverage maps corresponding to the two extreme positions of the transmitter node. However, the information provided in Fig. 4 is not enough to design a reliable receiver scheme, since it provides only macroscopic data about one point in time.

In this regard, Fig. 5 provides the information needed to evaluate the channel effects on the time dispersion of carrier assimilations, evaluated for a transmitter located close to the vessel walls ($d$=27.125µm).

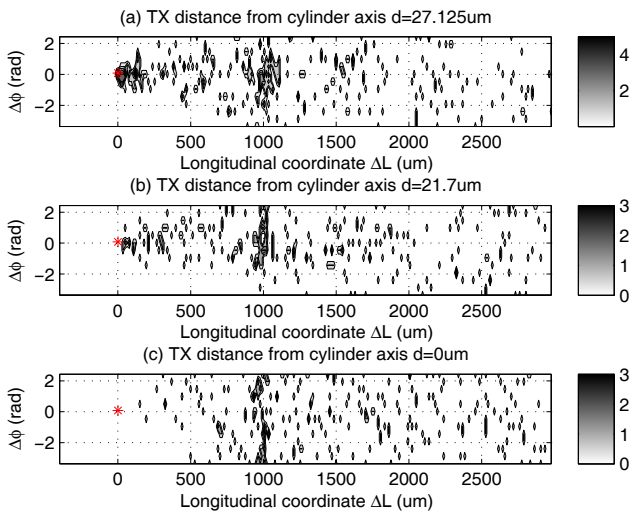

Fig. 4. Number of assimilated carriers after 8 s since the transmission.

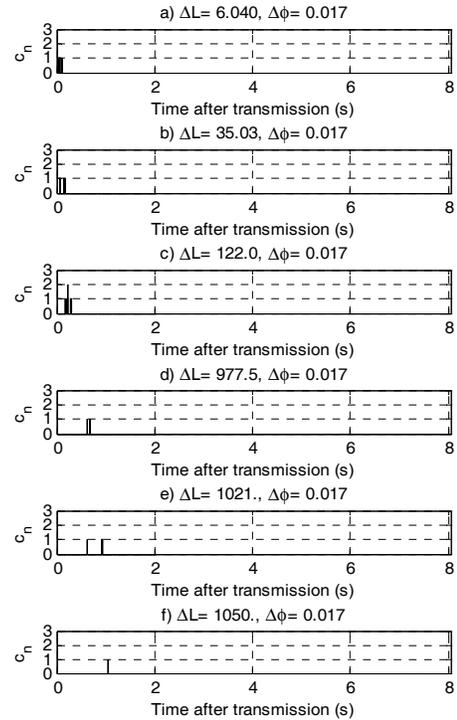

Fig. 5. Values of $c_n$ as a function of time since transmission, with $T$=750µs. Receivers are aligned with the transmitter ($\Delta\phi\approx$0 rad, see Fig. 1), and have a variable longitudinal displacement $\Delta L$, expressed in µm. At transmission time, the transmitter is in the closest position to the vessel wall (i.e., $d$=27.125µm).

Each subfigure in Fig. 5 shows the value of the output of the Counter(T) module ($c_n$) after the burst transmission. We have used a value of $T$=750µs, and taken samples from receivers aligned with the transmitter ($\Delta\phi\approx$0). This choice is motivated by the reception pattern shown in Fig. 4.(a).

When the distance $\Delta L$ of the receivers from the transmitter increases, we can observe a marked shift in time of the received signal. This is particularly evident in subfigure Fig. 5.(d), Fig. 5.(e), and Fig. 5.(f), which show the received signal at about 1 mm of distance from the transmission point. This effect is compliant with the value of the drag flow velocity close to the vessel walls, which is about half the mean flow velocity in the Poiseuille model, implemented in our simulator.

To go further, let us analyze the cumulative, long term number of assimilated carriers (that is the summation of $c_n$) shown in Fig. 4, in order to better identify possible values for the threshold $Th$ and the number of slots $P$, which determines the time window length. The first data can be derived from the analysis of Fig. 4. Since the maximum value of assimilated carriers is equal to 5 in the case of $d$<27.125 µm, i.e. transmitter very close to the vessel wall, it is reasonable to set the detection threshold to a value $Th \leq$5. In addition, observing also the reception patterns associated to larger distances from the vessel walls, we can conclude that it is necessary to consider a $Th \leq$3, since this value is the maximum number of assimilated carriers associated to position of the transmitter with $d$<27.125 µm. Thus, a value of $Th$ =2 seems reasonable in order to tolerate some variability in the distance of the transmitter from the vessel axis. As for the value of *P*, by looking at Fig. 5 we can deduce that the channel delay spread for fixed receivers, which are sticky to the vessel walls, is limited to few ms. The optimal value of *P* has been obtained by a brute force numerical analysis of the simulation results. The value of *P* which maximizes the number of cells able to receive the signal is $P_{opt}$=33 intervals. It corresponds to a filter time window to generate $f_n$ of about 25 ms, whereas the threshold value is $Th_{opt}$=2. Subfigures (a) to (d) of Fig. 6 shows all combinations of $P_{opt}$, $P_{opt} \times 16$, $Th_{opt}$, $Th_{opt} \times 2$, by using a burst of 3000 carriers. The size of this burst is compliant with the assumptions of [4], which consider each nanomachine with an initial stock of $10^8$ molecules, of which $10^3$–$10^4$ are emitted during each burst to transmit an information symbol [23]. We can see that using short values of both time window and threshold, it is possible to reach a large number of receivers, up to 1 mm from the transmission point (see also Fig. 5.(d) and Fig. 5.(e)). Instead, by enlarging the time window, the effect is that some receivers close to the transmitter will erroneously decode the single transmitted pulse as two different pulses. This effect is due to the so called channel memory [24]. Finally, using a larger *Th* value, independently of the time window, all receivers able to decode the signal will decode a single pulse. However, their number dramatically decrease, since the selected *Th* is too close to the maximum value of the assimilated carriers, and, in this case, increasing the time window does not produces benefits or drawbacks.

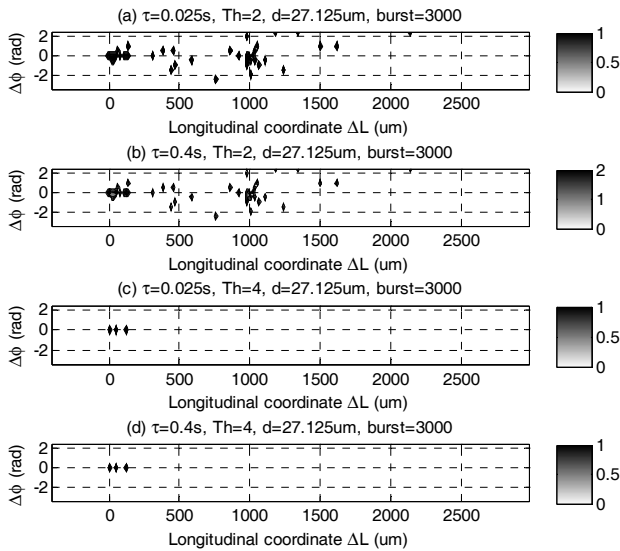

Fig. 6. Map indicating the receivers able to decode the impulse transmitted in the position (0,0). The side bar shows a chromatic scale to identify how many impulses have been decoded by appling the proposed decoding scheme in a simulation time equal to 8 seconds. Transmitter is located at *d*=27.125μm from the vessel axis. Burst of 3000 carriers.

## V. CONCLUSION

In this paper, we have proposed a decoding scheme for molecular communications in blood vessels. The preliminary performance figures shown are encouraging, and will be followed by a deeper analysis in order to provide an analytical model of the channel effects, which would implies more effective receivers. Future works will also consider simulations involving also mobile receivers.